\documentclass[natbib]{emulateapj}

\begin{document}

\title{Beyond Inside-Out Growth: Formation and Evolution of Disk Outskirts}

\author{Rok Ro\v{s}kar\altaffilmark{1}, Victor
P. Debattista\altaffilmark{1,2,3}, Gregory S. Stinson\altaffilmark{1,5},
Thomas R. Quinn\altaffilmark{1}, Tobias Kaufmann\altaffilmark{4},
James Wadsley\altaffilmark{5}}

\altaffiltext{1}{Astronomy Department, University of Washington, Box
351580, Seattle, WA 98195, USA {\tt
roskar;debattis;stinson;trq@astro.washington.edu}}
\altaffiltext{2}{Brooks Prize Fellow}
\altaffiltext{3}{RCUK Fellow at Centre for Astrophysics, University of Central
Lancashire, Preston, PR1 2HE, UK}
\altaffiltext{4}{Department of Physics and Astronomy, Centre for Cosmology,
University of California, Irvine, CA 92697 {\tt tobias.kaufmann@uci.edu}}
\altaffiltext{5}{Department of Physics and Astronomy, McMaster University, 
Hamilton, ON, L8S 4M1, Canada {\tt wadsley@mcmaster.ca}}

\begin{abstract}

We have performed a high mass and force resolution simulation of an
idealized galaxy forming from dissipational collapse of gas embedded
in a spherical dark matter halo. The simulation includes star
formation and effects of stellar feedback. In our simulation a stellar
disk forms with a surface density profile consisting of an inner
exponential breaking to a steeper outer exponential. The break forms
early on and persists throughout the evolution, moving outward as
more gas is able to cool and add mass to the disk. The parameters of
the break are in excellent agreement with observations.  The break
corresponds with a rapid drop in the star formation rate associated
with a drop in the cooled gas surface density, but the outer
exponential is populated by stars that were scattered outward on nearly circular orbits 
from the inner disk by spiral arms. The resulting profile and its associated
break are therefore a consequence of the interplay between a radial
star formation cutoff and redistribution of stellar mass by secular
processes. A consequence of such evolution is a sharp change in
the radial mean stellar age profile at the break radius.

\end{abstract}



\section{Introduction}
\label{sec:intro}

Since the early work of \citet{de-Vaucouleurs:1958} it
has been recognized that the disks of spiral
galaxies generally follow an exponential radial surface brightness
profile, and various theories have explored the physical
causes and consequences of this property \citep[e.g.][]{Fall:1980lr,
Lin:1987pb, Dalcanton:1997bh, Mo:1998mi,
van-den-Bosch:2001aa}. However, since \citet{van-der-Kruit:1979gb,
van-der-Kruit:1987fk} it has been known that the outer regions of
disks exhibit more varied behavior. This has been confirmed by an
abundance of recent data \citep[e.g.,][hereafter PT06]{Pohlen:2000ff,
Pohlen:2002ec, Bland-Hawthorn:2005ys, Erwin:2005kl, Pohlen:2006lh}.
In a sample of nearby late-type galaxies from the Sloan Digital Sky Survey,
PT06 found that about 60\% have an inner exponential followed by a steeper outer
exponential (downward-bending), $\sim$30\% have the inner exponential followed by a
shallower outer exponential (upward-bending), while only $\sim10$\% have no detectable
breaks.  Therefore breaks are a common feature of disk galaxies that
any complete theory of galaxy formation must be able to
explain. Furthermore, the discovery of UV emission at radii well
beyond the H$\alpha$ emission cutoff \citep{Gil-de-Paz:2005, Thilker:2005,
Thilker:2007a}, the observational evidence for inside-out disk
growth \citep{Munoz-Mateos:2007}, and detections of disk breaks in the
distant universe \citep{Perez:2004, Trujillo:2005}, suggest
that the outer disks provide a direct view of disk assembly.

Several theories for the formation of breaks have been investigated. 
\Citet{van-der-Kruit:1987fk} proposed that angular
momentum conservation in a collapsing, uniformly rotating cloud
naturally gives rise to disk breaks at roughly 4.5 scale radii.
\Citet{van-den-Bosch:2001aa} suggested that breaks are due to angular
momentum cutoffs of the cooled gas.  More commonly breaks have been
attributed to a threshold for star formation (SF), whether due to low
gas density which stabilizes the disk \citep{Kennicutt:1989bs}, or to
a lack of a cool equilibrium ISM phase \citep{Elmegreen:1994eu,
Schaye:2004kb}. Using a semi-analytic model, \citet{Elmegreen:2006oq} 
demonstrate that a double-exponential profile may result from a multi-component star
formation prescription.  The existence of extended UV disks
\citep[e.g][]{Thilker:2007a} and the lack of a clear
correlation of H$\alpha$ cut-offs and optical disk breaks
\citep{Pohlen:2004, Hunter:2006aa} further complicates the
picture. Regardless, while a sharp SF cutoff may explain the disk
truncation, it does not provide a compelling explanation for extended
outer exponential components.  Alternatively,
\citet{Debattista:2006wd} demonstrated that the redistribution of
angular momentum by spirals during bar formation also produces
realistic breaks in collisionless $N$-body simulations.

In this Letter we present the first results from a series of
high-resolution Smooth Particle Hydrodynamics (SPH) simulations of
isolated galaxy formation aimed at exploring the formation and
evolution of disk breaks and outskirts in a massive, high surface brightness galaxy
without a strong central bar.  
Resulting breaks are analogous to 
downward-bending breaks seen in observations. The clear advantage of our
approach over past attempts is that we use a fully self-consistent
physical model of the system, making no a priori assumptions about
the distribution of material in the disk.  Rather, we allow the disk
to grow spontaneously under the effects of gravity and gas
hydrodynamics, itself influenced by star formation and feedback. The
$N$-body approach (at sufficiently high mass and force resolution)
ensures that we capture the dynamical processes contributing to disk
evolution. Furthermore, the inclusion of prescriptions for SF and
feedback allows us to make observational predictions across the break
region. We show that (1) the break forms rapidly ($\lesssim$ 1Gyr) and
persists throughout the evolution of the system, moving outward as the
disk mass grows; (2) the break is seeded by a sharp decrease in star
formation which is caused in our simulation by a rapid decrease in the surface density
of cool gas; (3) the outer disk is populated by stars that have
migrated, on nearly circular orbits, from the inner disk, and
consequently the break is associated with a sharp change in the radial mean stellar 
age profile; (4) break parameters agree with current observations.


\section{Simulation Methodology}
\label{sec:methods}

Resolved stellar population data in disk outskirts, which are now
becoming available \citep[e.g.][]{Ferguson:2006, Barker:2007a, de-Jong:2007},
provide strong constraints on theories of break formation.  Therefore,
the inclusion of SF and feedback is required to assess break
formation models.  For this reason we have run simulations of gas
cooling, collapsing, and forming stars inside a live dark matter halo
within which the gas is initially in hydrostatic equilibrium.  This
has the further advantage of making no assumptions about the angular
momentum distribution within the disk, which can strongly affect its
subsequent evolution \citep{Debattista:2006wd}.

We construct initial conditions as in \citet{Kaufmann:2007}. The
initial system consists of a virialized spherical NFW dark matter halo
\citep{Navarro:1997aa} and an embedded spherical hot baryonic
component containing 10\% of the total mass and following the same 
density distribution, which at the end of the simulation
yields a disk mass fraction of $\sim$ 5\%. The mass within the
virial radius is $10^{12}$~M$_\odot$.  A temperature gradient in the
gas component ensures an initial gas pressure equilibrium for an adiabatic equation of state.  Velocities
of gas particles are initialized according to a cosmologically-motivated specific angular
momentum distribution with $j\propto r$ and an overall spin parameter
$\lambda = (j/G) \sqrt{|E|/M^3} = 0.039$ \citep{Bullock:2001}.  
Each component is modeled by $10^6$ particles; 
the dark matter halo is composed of two shells,
with the inner halo of $9\times10^5$ particles (each of mass
$\sim1\times10^6$ M$_{\odot}$) extending to 200~kpc and an outer halo
of $1\times10^5$ particles (each of mass $\sim3.5\times10^6$
M$_{\odot}$) beyond.  All gas particles initially have a mass of
$1.4\times10^5$ M$_{\odot}$.  We use a softening length of 100~pc for
all dark matter particles and 50~pc for baryonic particles.  We adopt the best-fit 
values from \citet{Stinson:2006aa} for the parameters governing the physics of 
star formation and feedback. Our cooling prescriptions do not account for effects of UV
background or metal line cooling.
The global criteria for SF are that a gas
particle has to have $n>$ 0.1~cm$^{-3}$, $T<15000$~K and be a part of
a converging flow; efficiency of star formation is 0.05,
i.e. 5\% of gas eligible to form stars is converted into stars per dynamical time. Star
particles form with an initial mass of 1/3 gas particle mass, which at
our resolution corresponds to $4.6\times10^4$ M$_{\odot}$. A gas
particle may spawn multiple star particles but to avoid gas particles of unreasonably
small mass, the minimum gas particle mass is restricted to
1/5 of the initial mass.  The simulation is evolved with the parallel
SPH code \textsc{gasoline} \citep{Wadsley:2004mb} for 10~Gyr. 

We have also performed simulations with $10^5$ and $10^7$ particles
per component, thus bracketing our fiducial run. While the details
of the gas cooling are somewhat resolution-dependent \citep{Kaufmann:2006th} resulting
in morphological differences, we find convergence in the modeling of overall halo cooling, 
star formation and its dependence on gas surface density, and disk dynamics. 
We therefore consider the conclusions and predictions 
of this Letter numerically robust.  The $10^6$ particle resolution used 
here represents a compromise between adequate statistics for 
detailed analysis in the outer disk and computational cost for 10 Gyr of evolution.

Our simulation should be thought of as modeling disk formation and
evolution from the cooling of hot gas after the last gas-rich major
merger, as suggested by cosmological simulations \citep{Brook:2004,
Brook:2007}. Although we make use of simplifications such as an
initially spherical distribution of matter, lack of halo substructure, 
the ignoring of subsequent accretion of dark matter and baryons, and an initial gas 
density profile that mimics that of the
dark matter, studying the idealized isolated case allows us to analyze
in detail the important dynamical processes affecting the evolution of
a massive isolated disk. The lessons we learn from this idealized case
will later be applied to galaxies evolved in a full cosmological
context (R. Ro\v{s}kar et al. in preparation).


\section{Break Formation and the Outer Disk}
\label{sec:break}
\begin{figure} 
   \centering
   \plotone{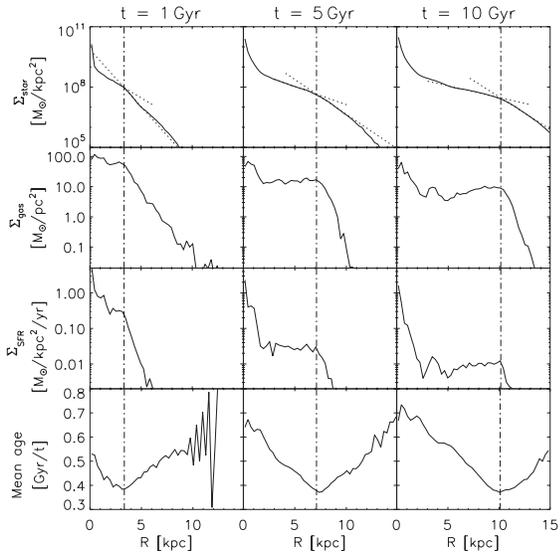}
   \caption{Azimuthaly-averaged properties of the disk 1, 5, and 10 Gyr.  
   The top panels show the
stellar surface density profiles, with the dotted lines representing
double exponential fits. The two exponential components were fit
simultaneously to the profile, excluding the innermost and outermost regions.
The point of intersection of the two
exponentials is taken as the break radius.  In all panels, the
vertical lines indicate the location of the break.  The second row
shows the surface density of cool gas. In the third row, we show the
SFR density calculated from the stellar mass formed in the previous 10
Myr.  The mean stellar age (normalized by time of output) as a
function of radius is shown in the bottom row.}
   \label{fig:dens_plots}
\end{figure}

In Figure~\ref{fig:dens_plots}, the break is already evident as soon as a stable disk forms at 1 Gyr,
moving outward as the disk grows and persisting throughout the
simulation. A sharp drop in the local SFR is always present at the
break radius. The drop in SFR is not due to a volume density threshold for star
formation, but is instead associated with a rapid decrease in
the gas surface density (the star
formation follows a Kennicutt-Schmidt law at all times) and a corresponding sharp increase
in the Toomre Q parameter. We verified that the break is not seeded by our star formation recipe
by running several simulations with different values of the threshold density and found that the 
location of the break did not depend on the particular value used. 
Since density is inversely proportional to radius, the cooling time increases outward. By construction,
the angular momentum is directly proportional to cylindrical radius, which means that higher 
angular momentum material will take longer to cool. This leads to the radial extent of the gas disk 
(and subsequently the star-forming disk) being limited by the maximum angular momentum of 
material that has been able to cool. 


\begin{figure}[tbp]
	\centering
	\plotone{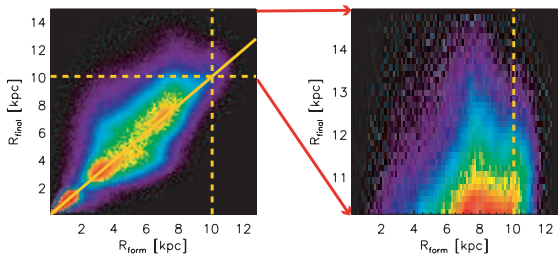}
	\caption{2D histogram of final particle radii vs. particle
	formation radii for particles on nearly circular orbits at 10 Gyr. 
	The right panel shows only particles beyond the break.}
	\label{fig:rformrfinal}
\end{figure}

Figure~\ref{fig:dens_plots} shows clearly that a sharp change in the mean stellar age
profile coincides with the location of the break. We now investigate the cause of this
ubiquitous feature.  Figure~\ref{fig:rformrfinal} shows a 2D
mass-weighted histogram of final ($t = 10$ Gyr) versus formation radii
of stellar particles on nearly circular orbits, defined as having $J_z/J_c(E) > 0.9$, 
where $J_c(E)$ is the angular momentum of a
circular orbit with the energy of a given particle.  This criterion is satisfied by
$\sim80\%$ of the outer disk particles. Therefore the outer disk is 
a rotationally-supported structure. It is clear from
Figure~\ref{fig:rformrfinal} that significant redistribution has
occurred throughout the disk; including all of the particles in this figure, the root-mean-square
change in radius during the lifetime of each particle $[\overline{(\Delta R)^2}]^{1/2} = 2.4$ kpc 
(mean $\Delta R \sim0$). Most strikingly, nearly
all ($\sim85$\%) particles beyond the break
(Figure~\ref{fig:rformrfinal}, \textit{right}) have migrated there from
the inner disk.  The mean epicycle radius of these \textit{outer disk} particles is only
$\sim2$ kpc, while the mean  $\Delta R = 3.7$ kpc.  Hence,
the presence in the outer disk of particles which formed in the inner
disk cannot be attributed to radial heating alone but is a signature of
secular radial redistribution. 

\citet{Sellwood:2002qf} describe how transient spiral arms move corotating stars radially
without heating the disk. Fourier analysis shows that spirals of a
wide range of pattern speeds are present in different parts of the
disk at all times, enabling shuffling of stars at all radii. 
As the disk size increases, spirals of ever larger corotation radii and lower pattern speeds
become common, enabling transport of material over large distances. 
The largest migrations coincide with the growth of strong spirals, 
confirming that the migrating stars are in resonance
with the spirals. Thus radial migration is common and stars of all
ages are continuously transported to well outside the break.  
Details of these mechanisms will be presented in a future paper.

Since the surface density break also corresponds to a break in SFR, the
outer disk is relatively deficient in young stars. The outermost
regions of the disk are populated by old stars which migrated from the
inner disk, since they have had the most time to reach there.  This
redistribution naturally leads to the minimum in stellar age at the
break and a positive age gradient in the outer disk. Although the importance
of spiral arms for the evolution of the outer disk has already been demonstrated
by \citet{Debattista:2006wd}, in their models the break was formed by 
resonant coupling between the central bar and spiral arms. The cause of the
break in our model is fundamentally different since it is ultimately caused by a drop in the 
SFR, while the outer exponential is populated by secularly redistributed stars. 
While a recurring oval does appear in the disk, a strong bar like the ones described
in \citet{Debattista:2006wd} does not form here.

The radial migrations are not directed only outward; indeed spirals
cause much more radial mixing than transport.
Figure~\ref{fig:rformrfinal} shows that the radial changes are
symmetric for particles with small deviations $\Delta R$ and at
smaller radii. However, for particles with larger $\Delta R$ or larger
initial radii, more particles move outward than inward. Part of the
excess angular momentum required for this transport is provided by the gas, which is funneled
toward the center.


\section{Comparison with Observations}
\label{sec:obs}

Our simulation is not
intended to reproduce all properties of real galaxies since we lack
the full cosmological context. Nevertheless, the evolution of disks in
isolated galaxies since the epoch of last major mergers at $z \sim 2$
(i.e. about 10 Gyr) should not be very different from that in our
model.  We may therefore anticipate that our model matches nature well
if it captures the essential physics.

We have investigated how the break structural parameters of our
simulated disk at different points in the simulation compare to those
observed by PT06 and \citet{Pohlen:2002ec}.  Ratios of break radius to
inner scalelength agree well with the data from PT06 throughout the simulation. 
The mean ratio of break radius to inner scalelength, $R_{br}/h_{in}$, in our
simulation is 2.6, while PT06 find a mean value of 2.5, an indication that 
our simulation captures the correct scale of the phenomenon.

\begin{figure}[tbp]
	\centering
	\plotone{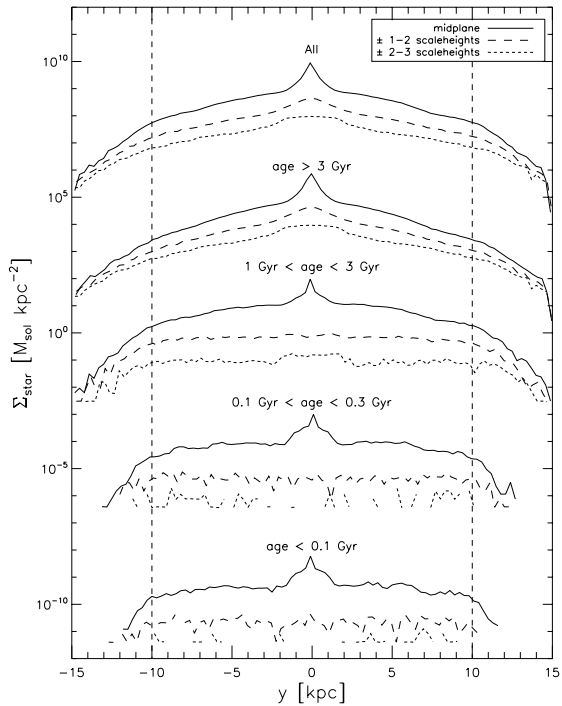}
	\caption{Stellar density profiles for stars in different age
	bins. An arbitrary offset has been applied for clarity. See
	text for details. }
	\label{fig:ageheight_prof}
\end{figure}

\citet{de-Jong:2007} presented radial star count profiles of the
edge-on galaxy NGC~4244 for different stellar populations. Their main result was that
the break occurs at the same radius regardless of age and distance
from the mid-plane. In Figure~\ref{fig:ageheight_prof} we show edge-on
density profiles for stars from our model with age bins roughly
corresponding to those used by \citet{de-Jong:2007}. Solid lines
correspond to stars in the midplane ($|z| < 0.5$ kpc), dashed lines
to stars at $0.5 < |z| < 1.0$ kpc, and dotted lines to stars at $1.0 < |z| < 1.5$ kpc. 
The vertical line marks the break
as determined from the face-on profile. The break occurs at the same
radius for all age bins, but becomes less pronounced with age due to
radial redistribution. In the two oldest age bins (where we have
enough particles to make a meaningful comparison), the break is also
constant with height away from the midplane. Transient spirals are responsible
for the largest amounts of redistribution, which is very efficient and can
operate on relatively short timescales.  
Therefore the agreement of the location of the break
between stars of different ages is a natural consequence of migration caused by
transient spirals in a growing disk. 

The decrease in mean stellar age with radius is unsurprising if disks
form inside-out, and such trends have indeed been observed
\citep{de-Jong:1996, Bell:2000, MacArthur:2004,
Munoz-Mateos:2007}. Thus, the negative age gradient out to the break is expected, but
the abrupt change in the gradient and its close correlation with the break radius has not
previously been noted. However, a positive age gradient beyond the break has been
observed in M33 \citep{Barker:2007a}, although, to the best of our knowledge, no
complementary age profile exists for the inner disk of that galaxy.

\section{Conclusions}
\label{sec:conclusions}

We have shown that in a self-consistent model, where the stellar disk
forms through gas cooling and subsequent star formation within a dark
matter halo, breaks in the stellar surface density form through the
combination of different effects. A rapid drop in the SFR seeds the
break and secular evolution populates the outer exponential.  In our
model the SFR drop is due wholly to a drop in the surface density of
gas.  However a break in SFR induced by other means ({\it e.g.} a
volume density threshold or perhaps warps) would lead to similar
behavior of the outer disk and stellar density break parameters.  
Our model properties satisfy current observational constraints, both in the statistical
sense of break properties from galaxies in SDSS,  as well as the much more detailed 
observations of breaks in stellar populations of NGC~4244.  
Though our model does not account for the
effects of evolution in a full cosmological context, its simplicity
assures that this is the minimal degree of evolution (with no
interactions or formation of a significant bar) and should therefore
be rather generic provided that a disk is massive enough for strong
transient spirals to form.  The model predicts that there should be an abrupt change
in the radial mean stellar age profile coincident with the break,
which can be tested with future observations.

\acknowledgments This research was supported in part by the NSF
through TeraGrid resources provided by TACC 
and PSC. Access to additional computing resources
was made possible by a grant from the U. W. Student Technology Fee and
support from the U. W. PACS team.  R. R., V. P. D., G. S. S., and T. R. Q. were supported by 
the NSF ITR grant PHY-0205413 at the University of Washington.
R. R. acknowledges support for a visit to the
University of Central Lancashire from a Livesey Award Grant held by
V. P. D.  Over the course of this paper V. P. D. was also supported by a Brooks
Prize Fellowship and an RCUK Fellowship at the University of Central Lancashire. 

\bibliographystyle{apj}

\clearpage

\end{document}